\documentclass{article}
\usepackage[breaklinks]{hyperref}
\usepackage[german,british]{babel}
\usepackage{fancyheadings}

\usepackage{graphicx}
\usepackage{psfig}
\usepackage{color}

\DeclareFontFamily{OT1}{rsfs}{}
\DeclareFontShape{OT1}{rsfs}{m}{n}{ <-7> rsfs5 <7-10> rsfs7 <10->rsfs10}{} 
\DeclareMathAlphabet{\mycal}{OT1}{rsfs}{m}{n}
\newcommand{\scri}{{\mycal I}}

\usepackage{amsmath}
\usepackage{amssymb}

\usepackage{theorem}
\newtheorem{theorem}{Theorem}
\theoremstyle{break}

\begin{document}

\title{Static perfect fluid balls with given equation of state and cosmological constant}
\author{Christian G. B\"ohmer\footnote{e-mail: {\tt boehmer@hep.itp.tuwien.ac.at}}} 
\date{}

\maketitle

%%% TUW preprint number on titlepage
\thispagestyle{fancy}
\setlength{\headrulewidth}{0pt}
\rhead{TUW--04--28}

\noindent
\textit{Institut f\"ur Theoretische Physik, Technische Universit\"at Wien,
Wiedner\\ Hauptstr. 8-10, A-1040 Wien, Austria\\ \mbox{}}

\begin{abstract}
Static and spherically symmetric perfect fluid solutions of Einstein's
field equations with cosmological constant are analysed. After showing
existence and uniqueness of a regular solution at the centre 
the extension of this solution is discussed.
Then the existence of global solutions with given equation of 
state and cosmological constant bounded by $4\pi\rho_b$, where 
$\rho_b$ is the boundary density (given by the equation of state)
of the perfect fluid ball, is proved.
\end{abstract}
\mbox{} \\ 
\mbox{} \\
\textit{Keywords: exact solutions, spherical symmetry, perfect fluid, 
cosmological\\ constant}
\mbox{} \\ 
\mbox{} 

\newpage
\section{Introduction}

This paper analyses static, spherically symmetric perfect 
fluid solutions to Einstein's field equations with cosmological 
constant for a given monotonic equation of state $\rho=\rho(P)$. 
The choice of central pressure (central energy density) and 
cosmological constant uniquely determines the pressure function. 
It is the aim of this work to extend the results presented 
in~\cite{Rendall:1991hg} to include the cosmological constant.

Existence and uniqueness of a global solution with given equation of state 
can be proved for cosmological constants satisfying 
$\Lambda < 4\pi \rho_{b}$, which is given by the equation
of state since $\rho_{b}=\rho(P=0)$, following the line of 
argument of~\cite{Rendall:1991hg}.

The existence of such global solutions is quite important because 
assuming their existence, Buchdahl~\cite{Buchdahl:1959} showed 
that the total mass of a fluid ball is bounded by its radius. 
He showed the strict inequality \mbox{$M < (4/9)R $}, which holds
for fluid balls in which the density does not increase outwards. 
It implies that radii of fluid balls are always larger than the 
black-hole event horizon. It is expected that perfect fluid balls
with cosmological constant are larger than the black hole event 
horizon but still be smaller than the cosmological event horizon.

This paper is organised as follows: In section~\ref{sec:exist}
Einstein's field equations and the Buchdahl variables are presented,
in the beginning of section~\ref{sec:existence} uniqueness and 
existence of a regular solution at the centre is shown. 
In theorem~\ref{th:extension} the extension of the solution 
is discussed. For fluid balls the Buchdahl inequality with 
cosmological term is obtained in theorem~\ref{th:buchdahl}. 
Solutions without singularities are constructed in 
section~\ref{sec:without}. Finally remarks on the 
finiteness of solutions are given and moreover some 
conclusions are presented.

\section{Field equations and Buchdahl variables}
\label{sec:exist}

The most general static and spherically symmetric
\begin{align}
      ds^{2}=-e^{\nu (r)}dt^{2}+e^{a(r)}dr^2
      +r^{2}(d\theta ^{2} +\sin^{2}\!\theta d\phi^{2})\,,
      \label{eq:metric}
\end{align} 
in Einstein's theory of gravity yields three independent field 
equations with cosmological constant
\begin{align}
      \frac{1}{r^{2}}e^{\nu(r)}\frac{d}{dr}\left(r-re^{-a(r)}\right) 
      -\Lambda e^{\nu(r)} &= 8\pi \rho(r)e^{\nu(r)}\,,
      \label{a} \\
      \frac{1}{r^{2}}\left(1+r\nu'(r)-e^{a(r)}\right) 
      +\Lambda e^{a(r)} &= 8\pi P(r)e^{a(r)}\,,
      \label{b} \\
      -\frac{\nu'(r)}{2}(P(r)+\rho(r)) &= P'(r)\,,
      \label{c} 
\end{align}
for an isotropic perfect fluid. Note that one can either use the three
field equations which imply conservation of energy-momentum or
two of the field equations together with the conservation equation 
which we do.

One function can be chosen freely since these are three independent 
ordinary differential equations for four unknown functions.
The most physical assumption is to prescribe an equation of state $\rho = \rho(P)$.
Integration of~(\ref{a}) gives 
\begin{align}
      e^{-a(r)}=1-2w(r)r^{2}-\frac{\Lambda}{3}r^{2}\,,
      \label{ea}
\end{align}
where the constant of integration is put to zero demanding regularity 
at the centre and where $w(r)$ is the mean density up to $r$ defined by
\begin{align}
	w(r)=\frac{m(r)}{r^{3}}\,, \qquad m(r)=4\pi\int_{0}^{r}s^2\rho(s)ds \,.
	\label{md}
\end{align}
Eliminating the function $\nu'(r)$ from the field equations
yields the Tolman-Oppenheimer-Volkoff~\cite{Oppenheimer:1939ne,Tolman:1939jz,Stuchlik:2000} 
equation (TOV-$\Lambda$)
\begin{align}
      P'(r)=-r\frac{\left(4\pi P(r) + w(r) -\frac{\Lambda }{3} \right) 
      \left(P(r)+ \rho(r) \right)}
      {1-2 w(r)r^{2}-\frac{\Lambda}{3}r^{2} }\,.
      \label{tov}
\end{align}

With given equation of state the conservation equation leads to
\begin{align}
      \nu(r)=-\int_{P_{c}}^{P(r)}\frac{2 d P}{P+\rho(P)}\,,
      \label{int}
\end{align}
where $P_{c}$ is the central pressure. 
With $m(r)$, equations~(\ref{md}) and~(\ref{tov})
are forming an integro-differential system for $\rho(r)$ and $P(r)$.
However, differentiating the mean density $w(r)$ with respect 
to $r$ implies
\begin{align}
      w'(r)=\frac{1}{r}\left(4\pi \rho(P(r)) - 3w(r) \right)\,.
      \label{mdd}
\end{align}
Therefore for a certain $\rho=\rho(P)$, equations~(\ref{tov}) and~(\ref{mdd}) 
are forming a system of first order differential equations in $P(r)$ and $w(r)$. 

To extract the TOV-$\Lambda$ equation from Einstein's field equations
the metric function $\nu(r)$ was eliminated. On the other hand,
one may eliminate the pressure for which Buchdahl~\cite{Buchdahl:1959} 
introduced new variables
\begin{align}
      y^{2} & = e^{-a(r)}=1-2w(r)r^{2}-\frac{\Lambda}{3}r^{2}\,,
      \label{eq:y} \\
      \zeta & = e^{\nu/2}\,, 
      \label{eq:zeta} \\
      x & =r^{2}\,,     	
      \label{eq:x}
\end{align}
which we have supplemented by a cosmological constant.
The second field equation (\ref{b}) written in those variables leads to
\begin{align}
      8\pi P-\frac{2}{3}\Lambda=4y^{2}\frac{\zeta_{,x}}{\zeta}-2w \,.
      \label{eq:PP}
\end{align}
After differentiating this with respect to $x$ and using 
the conservation equation (\ref{c}) 
the pressure can be eliminated. After some algebra one
arrives at
\begin{align}
      \left(y\zeta_{,x}\right)_{,x}
      -\frac{1}{2}\frac{w_{,x}\zeta}{y}=0 \,,
      \label{yzeta}
\end{align}
which is surprisingly similar to the square of the Weyl tensor
\begin{align}
      C_{abcd} C^{abcd} = \frac{64}{3}\frac{y^2}{\zeta^2} \Bigl( 
      (y\zeta_{,x})_{,x}+\frac{1}{2}\frac{w_{,x}\zeta}{y} 
      \Bigr)^2 x^2 \,.
      \label{eq:weyl}
\end{align}
This implies that the constant density solutions are conformally 
flat~\cite{Stephani:1967}, since $w_{,x}=0$ and therefore 
with~(\ref{yzeta}) gives $(y\zeta_{,x})_{,x}=0$.

\section{Existence of unique regular solutions}
\label{sec:existence}

Since the field equation for a static and spherically symmetric perfect
fluid with given equation of state reduce to a system of singular first order
differential equations it is our first aim to apply theorem~1 
of~\cite{Rendall:1991hg} to the resulting system 
with cosmological constant.

\begin{theorem}[Rendall and Schmidt, 1991] 
\label{th:exist}
Let $V$ be a finite dimensional real vector space, $ N:V \rightarrow V$ a 
linear mapping, $G:V \times I \rightarrow V$ a $C^{\infty}$ mapping and 
$g:I \rightarrow V$ a smooth mapping, where $I$ is an
open interval in $R$ containing zero. Consider the equation
\begin{align}
	s\frac{d f}{ds} + N f = s G(s,f(s))=g(s)\,,
	\label{eq:app1}
\end{align}
for a function $f$ defined on a neighbourhood of $0$ and $I$ and taking 
values in $V$. Suppose that each eigenvalue of $N$ has a positive real part. 
Then there exists an open Interval $J$ with $0 \in J \subset I$ and a 
unique bounded $C^{1}$ function $f$ on $J \setminus {0}$ satisfying (\ref{eq:app1}).
Moreover $f$ extends to a $C^{\infty}$ solution of (\ref{eq:app1}) on $J$ if $N,G$ 
and $g$ depend smoothly on a parameter $z$ and the eigenvalues of $N$ are 
distinct then the solution also depends smoothly on $z$.
\end{theorem}

Note that~(\ref{mdd}) is singular at the centre whereas~(\ref{tov}) 
is not, however using $\rho =\rho_{c} + x\rho_{1}$, where $\rho_{c}$ is the 
central density, we find for a given equation of state
\begin{align}
      x\frac{d w}{d x}+\frac{3}{2}w & =2\pi \rho_{c}+2\pi x\rho_{1}\,, 
      \label{rhs1} \\
      x\frac{d\rho_{1}}{d x}+\rho_{1} & =-\left(\frac{d P}{d\rho}\right)^{-1}
      \frac{1}{2}\frac{\left(4\pi P+ w -\frac{\Lambda}{3} \right)
      \left( P + \rho_{c} + x\rho_{1} \right)}
      {1-2w x-\frac{\Lambda}{3}x}\,.
      \label{rhs2}
\end{align}
With the help of the corresponding pressure relation 
$P=P_{c}+xP_{1}(\rho_{1})$ and by noting that
\begin{align}
      \left(1-2w x-\frac{\Lambda}{3}x\right)^{-1} =
      1+ &\left(2w x+\frac{\Lambda}{3}x\right) \nonumber \\
      &\times \left(1-2w x-\frac{\Lambda}{3}x\right)^{-1}\,,
\end{align}
we find that the matrix $N$ of~(\ref{eq:app1}) has the following form
\begin{align}
      \begin{pmatrix}
      3/2 & 0 \\
      \frac{(P_{c}+\rho_{c})}{2d P/d\rho (\rho_{c})} & 1
      \end{pmatrix}\,.
      \label{eq:N}
\end{align}
Since the eigenvalues of $N$ are independent of the cosmological 
constant the system has a unique bounded solution in the 
neighbourhood of the centre, which is indeed $C^{\infty}$. 
This implies the existence of a unique, smooth solution 
to~(\ref{tov}) and~(\ref{mdd}) or~(\ref{rhs1}) and~(\ref{rhs2}) near the centre. 

In~\cite{Collins:1983} it is claimed that for a fixed equation 
of state and cosmological constant the choice of central pressure 
(and by virtue of the equation of state the central density) do not 
uniquely determine the solution. The existence and
uniqueness theorem above disproves this statement.

Uniqueness of the solution at the centre immediately implies:
\begin{theorem}
Let an equation of state $\rho(P)$, a central pressure $P_{c}$ and the 
cosmological constant $\Lambda$ be given such that 
\begin{align}
      4\pi P_{c} + \frac{4\pi}{3}\rho(P_{c}) -\frac{\Lambda}{3} = 0\,,
      \label{eq:buch1}
\end{align}
where $(4\pi/3)\rho_c=(4\pi/3)\rho(P_{c})=w_{c}$ so that also
the central energy density is given by the equation of state.
Then the unique solution is the Einstein static universe 
with $\Lambda=\Lambda_{{\rm E}}$.
\end{theorem}

For well defined right-hand sides of~(\ref{rhs1}) and~(\ref{rhs2}) standard 
theorems for differential equations guarantee that the solution can be 
extended. This implies that the solution is extendible if the pressure 
is finite $P<\infty$ and if the denominator of~(\ref{rhs2}) is non-zero, 
i.e.~$y=1-2wx-(\Lambda/3)x > 0$. Since the second condition depends on
the cosmological constant, new properties may arise. Moreover it
must be clarified whether $y=0$ corresponds to a coordinate singularity
of the spacetime, as in the constant density case~\cite{Boehmer:2003uz},
or to a geometric singularity. In Buchdahl variables~(\ref{eq:y})--(\ref{eq:x}) 
the line element~(\ref{eq:metric}) takes the form
\begin{align}
      ds^{2}=-\zeta(x)^{2} dt^{2} + \frac{dx^2}{4x y(x)^{2}}+
      x(d\theta ^{2} +\sin^{2}\!\theta d\phi^{2})\,,
\end{align}
which implies the following non-vanishing components of
the Riemann tensor
\begin{align}
      R_{\theta t}{}^{\theta t} &= R_{\phi t}{}^{\phi t} =
      -2y(x)^{2}\,\frac{\zeta'(x)}{\zeta(x)}\,, \\
      R_{x \theta}{}^{x \theta} &= R_{x \phi}{}^{x \phi} =
      -2y(x)y'(x) \,, 
\end{align}
with the remaining two given by
\begin{align}      
      R_{\theta \phi}{}^{\theta \phi} = &\,\frac{1-y(x)^2}{x}\,, \\ 
      R_{xt}{}^{xt} = &-4xy(x)^2\,\frac{\zeta''(x)}{\zeta(x)} 
      - 2y(x)^2\,\frac{\zeta'(x)}{\zeta(x)} \nonumber \\
      &- 4xy(x)y'(x)\frac{\zeta'(x)}{\zeta(x)}\,,
\end{align}
which indicates that $y\rightarrow 0$ corresponds to a coordinate
singularity rather then a geometric singularity if $y'(x)$ 
and $1/\zeta(x)$ behave well as $y \rightarrow 0$.
\\ \mbox{} \\
The following theorem clarifies the extendibility of solutions.

\begin{theorem}
\label{th:extension}
Suppose the pressure is decreasing near the centre, which means
\begin{align}
	4\pi P_{c}+\frac{4\pi}{3}\rho(P_{c})-\frac{\Lambda}{3} > 0 \,.
	\label{eq:buch2}
\end{align}
Then the solution is extendible and the pressure is 
monotonically decreasing if
\begin{align}
	4\pi P+w-\Lambda/3 > 0\,.
        \label{eq:buch3}
\end{align}
\end{theorem}
\textbf{Proof.} Assume that $\rho=\rho(P)$, $P_{c}$ and $\Lambda$ are 
given such that $P$ is decreasing near the centre, then $w(x)_{,x} \leq 0$.
Since $y > 0$ equation~(\ref{yzeta}) implies
\begin{align}
	\left(y\zeta_{,x}\right)_{,x} \leq 0\,.
	\label{eq:buch4}
\end{align}
Rewriting~(\ref{eq:PP}) gives
\begin{align}
	y\zeta_{,x}=\frac{\zeta}{2y}\left(4\pi P+w-\frac{\Lambda}{3}\right),
	\label{eq:buch5}
\end{align}
next using the implication of~(\ref{eq:buch4}) leads to
\begin{align}
	y\zeta_{,x} \leq \left(y\zeta_{,x}\right)(0).
        \label{eq:buch6}
\end{align}
Together with the explicit expression of $y \zeta_{,x}$ in~(\ref{eq:buch5}) 
this finally shows
\begin{align}
	y \geq \frac{4\pi P+w-\frac{\Lambda}{3}}{4\pi P_{c}+w_{c}
		-\frac{\Lambda}{3}}.
	\label{eq:buch7}
\end{align}
Therefore the Buchdahl variable $y$ cannot vanish before the numerator does
and consequently the right-hand sides of (\ref{rhs1}) and (\ref{rhs2}) 
are well defined, hence one can extend the solution if $4\pi P+w-\Lambda/3 > 0$.

Since $y > 0$ and $4\pi P+w-\Lambda/3 > 0$ the sign of the 
right-hand side of (\ref{rhs2}) is strictly negative. Therefore the 
energy density and because of the equation of state the pressure 
are monotonically decreasing functions.\hfill $\Box$
\\ \mbox{} \\
The last theorem contained the extendibility condition of solutions,
which also allows to show the existence of global solutions.

\begin{theorem}
\label{th:main}
Suppose an equation of state is given such that $\rho$ is defined for $p \geq 0$,
non-negative and continuous for $p \geq 0$, $C^{\infty}$ for $p > 0$ and 
suppose that $d\rho/dp >0$ for $p > 0$. Furthermore assume that 
the cosmological constant be given such that $\Lambda < 4\pi \rho_{b}$.\footnote{Assumptions on the equation of state could be weakened according to~\cite{Baumgarte:1993} and~\cite{Mars:1996gd}. Moreover the line of argument presented in~\cite{Mars:1996gd} can surely be applied for cosmological constants having the derived bound.}

Then the pressure is decreasing near the centre and there 
exists for any positive value of the central pressure $P_{c}$ 
a unique inextendible static and spherically symmetric solution 
of Einstein's field equations with cosmological 
constant having a perfect fluid source with equation 
of state $\rho(P)$.  

If $\Lambda \leq 0$ the matter either occupies the whole spacetime
with $\rho$ tending to $\rho_{\infty}$ as $r$ tends to infinity or
the matter has finite extent. In the second case a unique 
Schwarzschild-anti-de Sitter solution attached as an exterior field.

If the cosmological constant satisfies $0<\Lambda<4\pi\rho_{b}$,
the matter has always finite extent
and a unique Schwarzschild-de Sitter solution attached as
an exterior field. 
\end{theorem}
\textbf{Proof.} If the cosmological constant is given such that $\Lambda<4\pi\rho_{b}$ then
\begin{align*}
      0 < \frac{4\pi}{3} \rho_{b} -\frac{\Lambda}{3}
        < 4\pi P_{c} + \frac{4\pi}{3} \rho(P_{c}) - \frac{\Lambda}{3}\,,
\end{align*}
and the pressure is decreasing near the centre by~(\ref{eq:buch2}).

Since the pressure is decreasing near the centre the 
denominator of~(\ref{eq:buch7}) is some positive number. 
Furthermore one can estimate the numerator of~(\ref{eq:buch7}) by
\begin{align}
      y & \geq \frac{4\pi P+w-\frac{\Lambda}{3}}
                    {4\pi P_c +w_c -\frac{\Lambda}{3}}\,, 
      \nonumber \\
	& \geq \frac{w_{b}-\frac{\Lambda}{3}}
                    {4\pi P_c +w_c -\frac{\Lambda}{3}}\,, 
      \label{eqn_estimate_wb} \\
	& \geq \frac{\frac{4\pi}{3}\rho_{b}-\frac{\Lambda}{3}}
                    {4\pi P_c +w_c -\frac{\Lambda}{3}}\,, 
      \label{eqn_estimate_rhob}
\end{align}
and conclude that if 
\begin{align}
      \Lambda < 4\pi \rho_{b}\,,
      \label{upper_limit}
\end{align}
the Buchdahl variable $y$ cannot vanish before the pressure does.
The coordinate $x_{b}$ where the pressure vanishes will be taken as the 
definition of the stellar object's radius $R$.

\subsection*{$\Lambda \leq 0$}
If $\Lambda \leq 0$ the matter can occupy the whole space 
because~(\ref{eq:buch7}) implies positivity of $y$. 

Suppose that $P(x_{b})=0$, at the corresponding radius $R$ the 
Schwarzschild-anti-de Sitter solution is joined uniquely by the 
condition $M=m(R)$. In this manner the metric is $C^{0}$ only, 
because the density at the boundary may be non-zero. 
The metric is $C^{1}$ at $P(R)=0$ if Gauss coordinates relative
to the hypersurface $P(R)=0$ are used. If the boundary 
density does not vanish the Ricci tensor has a discontinuity. 
Thus the metric is at most $C^{1}$. Without further
assumptions on the boundary density this cannot be improved.    

Now assume that $P(x) > 0$ for all $x>0$. $P(x)$ is monotonically decreasing, 
therefore $\lim_{x \rightarrow \infty} P(x) = P_{\infty}$ exists. This 
implies that the pressure gradient tends to zero as $x \rightarrow \infty$. 
Because $y^{-1} \rightarrow 0$ as $x \rightarrow \infty$ equation~(\ref{tov}) 
does not imply that $P_{\infty}=0$, which it does if $\Lambda=0$. 
Thus the equation of state only gives $\rho \rightarrow \rho_{\infty}=\rho(P_{\infty})$ as $x \rightarrow \infty$.

\subsection*{$0< \Lambda < 4\pi \rho_{b}$}
If $0< \Lambda < 4\pi \rho_{b}$ then one can estimate the 
pressure at the possible coordinate or geometric singularity 
$\hat{r}$ defined by $y(\hat{r})=0$ since
\begin{align}
	\frac{\Lambda}{3} < \frac{4\pi}{3} \rho_{b} \leq w_{b} \leq w(r)\,,
	\label{eqn_lambda3<wb}
\end{align}
which holds for all $r$. Therefore
\begin{align}
	P(\hat{r}) = \frac{1}{4\pi} \left( \frac{\Lambda}{3} - w(\hat{r}) \right) < 0\,.
\end{align}
Hence there exists $R$ such that $P(R)=0$. Since the
pressure is decreasing and $P(\hat{r})<0$ it follows that $R < \hat{r}$.  

Thus if the cosmological constant is positive and \mbox{$\Lambda < 4\pi \rho_{b}$} 
then the pressure always vanishes at some $x_{b}$. At the corresponding radius 
$R$ the Schwarzschild-de~Sitter solution is joined uniquely by the same 
condition $M=m(R)$. The metric is at most $C^{1}$ because the 
boundary density is larger than zero, this cannot be improved.\hfill $\Box$
\\ \mbox{} \\
It is quite remarkable that this upper bound $\Lambda<4\pi\rho_b$
was found independently by considering consistency of the Newtonian
limit~\cite{Nowakowski:2000dr}, the gravitational equilibrium
via the virial theorem~\cite{Nowakowski:2001zw} and was also found
in~\cite{Sussman:2003km} by demanding stability of circular orbits.
This ``coincidence'' deserves a closer inspection, which was 
already initiated in~\cite{Balaguera-Antolinez:2004sg}.

\section{Buchdahl inequality}

The importance of Buchdahl's inequality was already discussed in
the introduction. In the following its generalisation with
cosmological constant is derived, see e.g.~\cite{Beig:2000yf}
without the cosmological term and~\cite{Mak:2001gg} with
cosmological constant following Buchdahl's original approach.

In the proof of the following theorem the field equations 
with constant density, denoted with a tilde, 
are solved with the help of Buchdahl variables
and then this solution is compared with a general decreasing solution.
The boundary mean density of the general solution defines the constant
density solution that is used for comparison.

\begin{theorem}
\label{th:buchdahl}
Let the cosmological constant be given such that $\Lambda < 4\pi \rho_{b}$.
Then for solutions having finite radius there holds
\begin{align}
      \sqrt{1-2w_{b} R^{2}-\frac{\Lambda}{3}R^{2}}  
      \geq \frac{1}{3} - \frac{\Lambda}{9 w_{b}}\,.
\end{align}
\end{theorem}
\emph{Proof.} Assume that $\rho(P)$, $P_{c}$ and 
$\Lambda$ are given such that the pressure is decreasing
near the centre. Then by theorem~\ref{th:extension} the pressure 
and mean density are decreasing functions and equation~(\ref{yzeta}) implies 
\begin{align}
	(\tilde{y}\tilde{\zeta}_{,x})_{,x}=0\,, \qquad
        \tilde{y}\tilde{\zeta}_{,x}=\frac{1}{2}\tilde{w}-\frac{\Lambda}{6}\,,
	\label{zetatilde}
\end{align}
where the constant of integration is obtained 
from~(\ref{eq:PP}) evaluated at the boundary
which can be used to integrate (\ref{zetatilde}). 
Let us rewrite the right-hand side of (\ref{zetatilde}) as
\begin{align}
	\tilde{y}\tilde{\zeta}_{,x}=
	\left( 2\tilde{w}+\frac{\Lambda}{3} \right)
	\left( \frac{1}{2}\tilde{w}-\frac{\Lambda}{6} \right)
	\left( 2\tilde{w}+\frac{\Lambda}{3} \right)^{-1}\,,
\end{align}
and substitute in $-2\tilde{y}\tilde{y}_{,x}=2\tilde{w}+\Lambda/3$ for the first 
factor, divide by $\tilde{y}$ which after integration yields 
\begin{align}
      \tilde{\zeta}(x)=
      -\frac{\tilde{w}-\frac{\Lambda}{3}}{2\tilde{w}+\frac{\Lambda}{3}}\tilde{y}(x)
      +\tilde{\zeta}(0)+\frac{\tilde{w}
      -\frac{\Lambda}{3}}{2\tilde{w}+\frac{\Lambda}{3}}\,.
\end{align}
From equation~(\ref{eq:buch4}) we conclude 
$\tilde{y}\tilde{\zeta}_{,x} = \left(y\zeta_{,x}\right)_{b} < y\zeta_{,x}$,
so that with $\tilde{y} > y$ one finds
\begin{align}
	\zeta(x) \geq \tilde{\zeta}(x) = 
        -\frac{\tilde{w}-\frac{\Lambda}{3}}{2\tilde{w}+\frac{\Lambda}{3}}
	\tilde{y}(x)+\tilde{\zeta}(0)+\frac{\tilde{w}-\frac{\Lambda}{3}}
        {2\tilde{w}+\frac{\Lambda}{3}}\,.
\end{align}
Evaluating this at the boundary and using that $\zeta$ can be 
normalised such that $\zeta_b =y_b$, it is found that
\begin{align}
	y_{b}  \geq  -\frac{\tilde{w}-\frac{\Lambda}{3}}{2\tilde{w}+\frac{\Lambda}{3}}
	y_{b} +\tilde{\zeta}(0)+ \frac{\tilde{w}-\frac{\Lambda}{3}}{2\tilde{w}+\frac{\Lambda}{3}}\,.
\end{align}
Since $\tilde{\zeta}(0)$ is a positive number, some algebra
finally leads to the generalised Buchdahl inequality 
\begin{align}
	y_{b} \geq \frac{1}{3} - \frac{\Lambda}{9 \tilde{w}}\,.
	\label{buchdahlw}
\end{align}
Now we compare solutions with decreasing mean density and solution
with constant density, where the constant density corresponds 
to the boundary mean density of the general solution. 
Thus~(\ref{buchdahlw}) yields
\begin{align}
	y_{b} \geq \frac{1}{3} - \frac{\Lambda}{9 w_{b}}\,,
	\label{buchdahlrho}
\end{align}
which reads more explicitly
\begin{align}
	\sqrt{1-2w_{b} R^{2}-\frac{\Lambda}{3}R^{2}}  
	\geq \frac{1}{3} - \frac{\Lambda}{9 w_{b}}\,,
	\label{eqn_buch_sqr}
\end{align}
and which holds for all monotonically decreasing densities.\hfill $\Box$
\\ \mbox{} \\
By using the definition of the boundary mean density $w_b = M/R^3$ one
can reformulate~(\ref{eqn_buch_sqr}) arrive at
\begin{align}
      3M < \frac{2}{3} R + R \sqrt{\frac{4}{9}-\frac{\Lambda}{3}R^2}\,,
      \label{eq:buch10}
\end{align}
a generalised Buchdahl inequality that takes the cosmological constant
into account. In~\cite{Mak:2001gg} the $2M/R$ version of~(\ref{eq:buch10})
was for example used to calculate the surface redshift with $\Lambda$.
For comparison, note the in~\cite{Mak:2001gg} the cosmological constant
is rescaled by $8\pi$.

\section{Solutions without singularities}
\label{sec:without}

The last two sections showed the existence of stellar models for
cosmological constants $\Lambda < 4\pi \rho_{b}$. 
Equation~(\ref{eq:buch10}) implies that the boundary of the stellar object 
has a lower bound given by the black-hole event horizon 
and an upper bound given by the cosmological event horizon. 
The upper bound for positive cosmological constants.

Stellar models with $\Lambda \leq 0$ have a lower bound
given by the black-hole event horizon. At the boundary
the Schwarzschild-anti-de Sitter solution attached as
an exterior field. Therefore
stellar models with $\Lambda \leq 0$ have no singularities.
Solutions with cosmological constant satisfying $\Lambda \leq 0$  
are globally static.

For positive cosmological constants the situation is different. But 
one can also construct solutions without singularities.

At the boundary $r=R$, defined by $P(R)=0$ the Schwarzschild-de Sitter solution
is joined $C^{1}$ by the usual procedure introducing Gauss coordinates.
The surface $r=R$ can also be found in the vacuum region where
the time-like Killing vector is past directed. This means that a second
stellar object can be put in the spacetime:

\begin{figure}[ht!]
\begin{center}
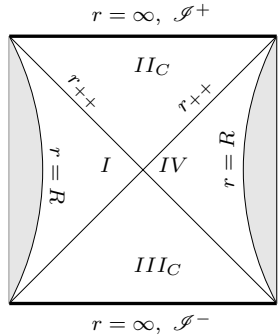
\end{center}
\caption{Penrose-Carter diagram with two stellar objects having radii $R$ which lie between
        the two horizons. Since the group orbits are increasing up to $R$ the vacuum
        part contains the cosmological event horizon $r_{++}$.
        This solution with two objects has no singularities. Because of regions $II_{C}$
        and $III_{C}$ this spacetime is not globally static.}
\label{fig: Penrose2}
\end{figure}

\subsection*{Remarks on finiteness of the radius}

So far it has been shown that given an equation of state, a central 
pressure and a cosmological constant there exist a unique model 
with finite or infinite extent. This depends on the given
equation of state and on the cosmological constant. As was
proved in theorem~\ref{th:main}, solutions are always finite
for positive cosmological constants.
If $\Lambda \leq 0$ either the pressure vanishes for some finite 
radius or the density is always positive and tends to $\rho_{\infty}$ 
as $r$ tends to infinity, which depends on the equation of state;
in~\cite{Rendall:1991hg} necessary and sufficient conditions can be found.

\section{Conclusions and Outlook}

In this paper existence and uniqueness of static and spherically
symmetric perfect fluid solutions with given equation of state
for cosmological constants satisfying $\Lambda <4\pi\rho_b$
was proven. It would be interesting to extend these proofs to
any value of the cosmological constant. The existence of these
solutions is conjectured in~\cite{Balaguera-Antolinez:2004sg}. 

One could start studying this system numerically or by using generating
function techniques. An investigation of the Riemann tensor
as outlined might also be a good starting point.

Furthermore the relation of the different theories that give
rise to the same upper bound must be analysed and understood.

\section*{Acknowledgements}
This article is based on the author's unpublished diploma thesis
which was supervised by Bernd G. Schmidt.
I am grateful to thank Herbert Balasin, Daniel Grumiller and Wolfgang Kummer 
for their constant support and moreover I thank Marek
Nowakowski for his suggestions regarding this project.

This work has partly been supported by project P-15449-N08 of 
the Austrian Science Foundation (FWF) and is part of the research 
project Nr. 01/04 {\it Quantum Gravity, Cosmology and Categorification} 
of the Austrian Academy of Sciences (\"OAW) and the National
Academy of Sciences of Ukraine (NASU).

\addcontentsline{toc}{section}{References}
\bibliographystyle{plain}
\bibliography{/home/boehmer/spinors/SRG/add_ref}

\begin{thebibliography}{10}

\bibitem{Balaguera-Antolinez:2004sg}
A.~Balaguera-Antol{\'{\i}}nez, C.~G. {B\"ohmer}, and M.~Nowakowski.
\newblock On astrophysical bounds of the cosmological constant.
\newblock 2004.
\newblock {\it gr-qc/0409004}.

\bibitem{Baumgarte:1993}
T.~W. Baumgarte and A.~D. Rendall.
\newblock Regularity of spherically symmetric static solutions of the
  {E}instein equations.
\newblock {\em Class. Quant. Grav.}, 10:327--332, 1993.

\bibitem{Beig:2000yf}
R.~Beig and B.~G. Schmidt.
\newblock Time-independent gravitational fields.
\newblock {\em Lect. Notes Phys.}, 540:325--372, 2000.

\bibitem{Boehmer:2003uz}
C.~G. B{\"o}hmer.
\newblock Eleven spherically symmetric constant density solutions with
  cosmological constant.
\newblock {\em Gen. Rel. Grav.}, 36:1039--1054, 2004.

\bibitem{Buchdahl:1959}
H.~A. Buchdahl.
\newblock General relativistic fluid spheres.
\newblock {\em Phys. Rev.}, 116:1027--1034, 1959.

\bibitem{Collins:1983}
C.~B. Collins.
\newblock Comments on the static spherically symmetric cosmologies of {E}llis,
  {M}aartens, and {N}el.
\newblock {\em J. Math. Phys.}, 24:215--219, 1983.

\bibitem{Mak:2001gg}
M.~K. Mak, Jr. Dobson, P.~N., and T.~Harko.
\newblock Maximum mass-radius ratio for compact general relativistic objects in
  {S}chwarzschild-de {S}itter geometry.
\newblock {\em Mod. Phys. Lett.}, A15:2153--2158, 2000.

\bibitem{Mars:1996gd}
M.~Mars, M.~M. Martin-Prats, and J.~M.~M. Senovilla.
\newblock The {$2m <= r$} property of spherically symmetric static spacetimes.
\newblock {\em Phys. Lett.}, A218:147--150, 1996.

\bibitem{Nowakowski:2000dr}
M.~Nowakowski.
\newblock {The consistent Newtonian limit of Einstein's gravity with a
  cosmological constant}.
\newblock {\em Int. J. Mod. Phys.}, D10:649--662, 2001.

\bibitem{Nowakowski:2001zw}
M.~Nowakowski, J.-C. Sanabria, and A.~Garcia.
\newblock {Gravitational equilibrium in the presence of a positive cosmological
  constant}.
\newblock {\em Phys. Rev.}, D66:023003, 2002.

\bibitem{Oppenheimer:1939ne}
J.~R. Oppenheimer and G.~M. Volkoff.
\newblock On massive neutron cores.
\newblock {\em Phys. Rev.}, 55:374--381, 1939.

\bibitem{Rendall:1991hg}
A.~D. Rendall and B.~G. Schmidt.
\newblock Existence and properties of spherically symmetric static fluid bodies
  with a given equation of state.
\newblock {\em Class. Quant. Grav.}, 8:985--1000, 1991.

\bibitem{Stephani:1967}
H.~Stephani.
\newblock {\"U}ber {L}{\"o}sungen der {E}insteinschen {F}eldgleichungene, die
  sich in einen f{\"u}nfdimensionalen flachen {R}aum einbetten lassen.
\newblock {\em Commun. Math. Phys.}, 4:137--142, 1967.

\bibitem{Stuchlik:2000}
Z.~Stuchl{\'{\i}}k.
\newblock Spherically symmetric static configurations of uniform density in
  spacetimes with a non-zero cosmological constant.
\newblock {\em Acta Physica Slovaca}, 50:219--228, 2000.

\bibitem{Sussman:2003km}
R.~A. Sussman and X.~Hernandez.
\newblock On the {N}ewtonian limit and cut--off scales of isothermal dark
  matter halos with cosmological constant.
\newblock {\em Mon. Not. Roy. Astron. Soc.}, 345:871, 2003.

\bibitem{Tolman:1939jz}
R.~C. Tolman.
\newblock Static solutions of {E}instein's field equations for spheres of
  fluid.
\newblock {\em Phys. Rev.}, 55:364--373, 1939.

\end{thebibliography}

\end{document}